\documentclass{amsart}
\newtheorem{theorem}{Theorem}[section]

\newtheorem{proposition}[theorem]{Proposition}
\theoremstyle{definition}
\newtheorem{definition}[theorem]{Definition}
\newtheorem{example}[theorem]{Example}

\theoremstyle{remark}

\numberwithin{equation}{section}
\newcommand{\abs}[1]{\lvert#1\rvert}

\newcommand{\N}{\mathcal{N}}
\newcommand{\C}{\mathbb{C}}
\newcommand{\WH}{\widehat{\mathfrak{H}}}
\newcommand{\IH}{\mathfrak{H}}
\newcommand{\KW}{\mathcal{W}}
\newcommand{\PA}{\mathfrak{A}}
\newcommand{\D}{\mathfrak{D}}
\newcommand{\W}{\mathcal{W}}

\begin{document}
\title{COHERENT STATES with  complex functions}
\author{K. Thirulogasanthar$^{\dagger}$}
\address{${}^{\dagger}$\;Department of Mathematics and Statistics, Concordia University,
   7141 Sherbrooke Street West, Montreal, Quebec H4B 1R6, Canada }
\email{santhar@vax2.concordia.ca}
\author{ G. Honnouvo$^{\ddagger}$}
\address{${}^{\ddagger}$\;Department of Mathematics and Statistics, Concordia University,
  7141 Sherbrooke Street West, Montreal, Quebec H4B 1R6, Canada.}
\address{${}^{\ddagger}$\;Unit\'e de Recherche en Physique Th\'eorique, Institut de Math\'ematiques et de Sciences Physiques, 01 B.P. 2628 Porto-Novo, Benin.\;International Chair in Mathematical Physics and Applications, 01 B.P. 2628 Porto-Novo, Benin.} 
\email{g$_{\_}$honnouvo@yahoo.fr}
\date{\today}
\begin{abstract}
The canonical coherent states are expressed as infinite series in powers of a complex number $z$ in their infinite series version. In this article we present classes of coherent states by replacing this complex number $z$ by  other choices, namely, iterates of a complex function, higher functions and elementary functions. Further, we show that some of these classes do not furnish generalized oscillator algebras in the natural way. A reproducing kernel Hilbert space is discussed to each class of coherent states.
\end{abstract}
\maketitle
\section{Introduction}
Hilbert spaces are the natural framework of mathematical aside of many branches of applications, for instant, quantum theories, signal and image analysis. As far as physical applications are concerned, the most fundamental for the analysis, or decomposition, of states in the Hilbert space of the problem is an over complete family of vectors known as coherent states (CS for short). In the terminology of signal processing, the Hilbert space consist of finite energy signals (square integrable functions), here again one intend to decompose a signal and select a suitable set of components in such a way that, in the reconstruction process one recover the signal without loosing essential information of the signal. In order to accomplish this task practitioners prefer to have an over complete family of vectors in the Hilbert space (for practical reasons an orthonormal basis is not a suitable one \cite{Du}). This family is generally known as frames. A most suitable family, in practice, is known as wavelets. In fact, wavelets are CS, namely, those associated to the affine group of appropriate dimension. Further, CS are a specific type of tight frames, for a detail description see \cite{key1}.
\bigskip

There are a number of ways to define a set of CS, for example, see \cite{key1, key2,key4} for different approaches. In this article we define it as a generalized version of the canonical CS (infinite series version). In fact, this type of CS were discussed in the literature in a somewhat different context, see theorem 5.4.2 of \cite{key1}.
\begin{definition}\label{D1}
Let $\mathfrak H$ be a separable Hilbert space with an orthonormal basis $\{\phi_{m}\}_{m=0}^{\infty}$ and $\mathbb C$ be the complex plane. Let $\D$ be an open subset of $\mathbb C$ and define
$$\Phi_m: \D\longrightarrow \D,\;\;\;m=0,1,2,\dots,$$
a sequence of complex functions. Then the vectors
\begin{equation}
\mid \Phi(z)\rangle=\mathcal N(|z|)^{-\frac{1}{2}}\sum_{m=0}^{\infty}\frac{\Phi_{m}(z)}{\sqrt{\rho(m)}}\phi_{m}\in\mathfrak{H};\;\;z\in\D
\label{1}\end{equation}
are said to form a set of CS if
\begin{enumerate}
\item[(a)]
For each $z\in\D$, the states $\mid \Phi(z)\rangle$ are normalized, that is, $\langle\Phi(z)\mid\Phi(z)\rangle=1.$
\item[(b)]
The states $\{\mid \Phi(z)\rangle\;:\;z\in\D\}$ satisfy a resolution of the identity, that is
\begin{equation}
\int_{\mathfrak D}\mid \Phi(z)\rangle \langle \Phi(z)\mid d\mu=I\label{2}
\end{equation}
\end{enumerate}
where $\mathcal N(|z|)$ is the normalization factor, $\{\rho(m)\}_{m=0}^{\infty}$ is a sequence of nonzero positive real numbers, $d\mu$ is an appropriately chosen measure and $I$ is the identity operator on $\mathfrak H$.\label{d1}
\end{definition}
The definition \ref{D1} is a generalization of the well-known classes of CS,
\begin{equation}\label{E}
\mid z\rangle=\mathcal N(|z|)^{-\frac{1}{2}}\sum_{m=0}^{\infty}\frac{z^m}{\sqrt{\rho(m)}}\phi_m
\end{equation}
where $z\in\mathbb \D$, an open subset of $\C$ and $\rho(m)$ is as in definition \ref{D1}. For $\rho(m)=m!$ we get $\N(|z|)=e^{|z|^2}$, this choice is the basic harmonic oscillator canonical CS.

\bigskip

For $\Phi_m(z)=z^m$ the states (\ref{1}), for various $\rho(m)$, were studied extensively and applied successfully in quantum theories for almost half a century. Interesting links were established with group representations classical polynomials etc. \cite{key1,key2,key7,key17,key4}.
Recently another class of CS were introduced, by Gazeau and Klauder, for Hamiltonians with discrete and continuous spectrum by taking $\Phi_m(z)=(\sqrt{J})^me^{ie_m\alpha}$ and $\rho(m)=e_1e_2...e_m$, where $e_m$'s are the spectrum of the Hamiltonian arranged in a particular way, for details see \cite{GK}. Following thir work  the same class of CS were studied  for various Hamiltonians, for example see \cite{Antoine}. 

\bigskip

Due to the wide range of applications of CS in quantum theories, it is an active area of research to find new classes of CS in various domains. In the literature, a huge volume of articles were devoted to this task. Main attention was paid on the construction of CS by changing $\rho(m)$ in (\ref{E}), for example see \cite{key2,key7}. However, in a recent article, \cite{key9} CS agreeing with definition (\ref{d1}) were presented by taking $\Phi_m(z)=Z^m$, where $Z$ is a $n\times n$ matrix valued function. It was also argued that these CS (in fact vector coherent states) are constructive in studying multi-level spin atoms in electromagnetic fields.

\bigskip

 In this article we present new classes of CS by replacing the function $\Phi_m(z)$ of definition \ref{D1} by the $m^{\text{th}}$ iterate of a complex function, Laguerre polynomials and Bessel function of the first kind. As the last part of this paper we discuss classes of CS obtained by replacing $\Phi_m(z)$ by some elementary functions.

\bigskip

For the states (\ref{E}) a naturally associated Lie algebra structure can be defined, see section \ref{osc}. However, we show that for the states (\ref{1}) it is not always possible to define such algebra. But the cases where $\Phi_m(z)=F(z)^m$, the $m^{\text{th}}$ power of some function, we can define such an algebra. We discuss such examples in section \ref{Eli}.

\bigskip

Mathematically our construction allow, through the resolution of the identity, one to decompose any vector of any separable Hilbert space in terms of $\Phi_m(z)$, which is a special function, iterate of a function, elementary functions (when one changes the Hilbert space, the basis elements $\phi_m$ changes accordingly). Further, as a consequence of the resolution of the identity, we will also associate a reproducing kernel and an isometric embedding between certain Hilbert spaces.
\section{CS labeled by the iterates of a complex function}
In this section we introduce a class of CS by taking $\Phi_m(z)$ as the $m^{\text{th}}$ iterate of a complex function $T$. We first discuss a general scheme to build CS together with a discussion of the construction of a reproducing kernel Hilbert space and then present an example.\\

Let
$T:\C\longrightarrow\C$
is a complex function. Let $T^{m}(z)=T\circ T\circ T\circ...\circ T(z)$ denote the $m^{\text{th}}$ iteration and $T^{0}(z)=z$.
\begin{theorem}\label{th1}
 Let $\{\phi_m\}_{m=0}^{\infty}$ be an orthonormal basis of an abstract separable Hilbert space $\IH$. Suppose the function $T$ has the following properties for some positive sequence $\{\rho(m)\}_{m=0}^{\infty}.$
\begin{eqnarray}
&&\mathcal N(|z|)=\sum_{m=0}^{\infty}\frac{|T^{m}(z)|^{2}}{\rho(m)}<\infty\label{3}\;\;\text{for each}\;z\in\C\\
&&\int_{\mathbb C}\frac{1}{\mathcal N(|z|)}T^{m}(z)\overline{T^{n}(z)}d\mu(z)=\left\{
\begin{array}{ccc}
0&\text{if}&m\not=n\\
\rho(m)&\text{if}&m=n
\end{array}\right.\label{4}
\end{eqnarray}
where $d\mu$ is an appropriately chosen measure. Then the vectors
\begin{equation}
\mid T(z)\rangle=\N(|z|)^{-\frac{1}{2}}\sum_{m=0}^{\infty}\frac{T^{m}(z)}{\sqrt{\rho(m)}}\phi_{m}\in\IH;\;\;z\in\C\label{5}
\end{equation}
form a set of CS.
\end{theorem}
\begin{proof}
We require $$\langle T(z)\mid T(z)\rangle=1.$$ This requirement can be obtained as follows:
\begin{eqnarray*}
\langle T(z)\mid T(z)\rangle&=&\mathcal N(|z|)^{-1}\sum_{m=0}^{\infty}\sum_{l=0}^{\infty}\frac{T^{m}(z)\overline{T^{l}(z)}}{\sqrt{\rho(m)\rho(l)}}\langle\phi_{m}\mid\phi_{l}\rangle\\
&=&\mathcal N(|z|)^{-1}\sum_{m=0}^{\infty}\frac{|T^{m}(z)|^{2}}{\rho(m)}=1,\\
\end{eqnarray*}
where we have used (\ref{3}). Thus, the states, $\mid T(z)\rangle$ are normalized. For a resolution of the identity, consider
\begin{eqnarray*}
\int_{\mathbb C}\mid T(z)\rangle\langle T(z)\mid d\mu(z) &=&\sum_{m=0}^{\infty}\sum_{l=0}^{\infty}\frac{1}{\sqrt{\rho(m)\rho(l)}}\int_{\mathbb C}\mathcal N(|z|)^{-1}T^{m}(z)\overline{T^{l}(z)}\mid \phi_{m}\rangle\langle\phi_{l}\mid d\mu(z)\\
&=&\sum_{m=0}^{\infty}\frac{1}{\rho(m)}\int_{\mathbb C}\mathcal N(|z|)^{-1}T^{m}(z)\overline{T^{m}(z)}d\mu(z)\mid \phi_{m}\rangle\langle\phi_{m}\mid \\
&=&\sum_{m=0}^{\infty}\mid \phi_{m}\rangle\langle\phi_{m}\mid=I
\end{eqnarray*}
where we have used (\ref{4}). We have obtained a resolution of the identity
\begin{equation}\label{res}
\int_{\mathbb C}\mid T(z)\rangle\langle T(z)\mid d\mu(z)=I
\end{equation}
 Thus the collection of vectors, (\ref{5}) forms a set of CS.
\end{proof}
Let $\WH=L^2(\C,d\mu)$ be the Hilbert space of all complex valued functions on $\C$ that are square integrable with respect to the measure $\mu$. From the resolution of the identity (\ref{res}), for any $\phi\in\IH$, we have
$$\phi=\int_{\C}\langle T(z)\mid \phi\rangle \mid T(z)\rangle d\mu$$
that is,
\begin{equation}\label{map1}
\Psi:\C\longrightarrow\C\;\;\;\text{with}\;\;\;\Psi(z)=\langle T(z)\mid \phi\rangle;\;\;\phi\in\IH
\end{equation}
defines elements in $\WH$. Further
\begin{equation}\label{map2}
\W:\IH\longrightarrow\WH\;\;\;\text{with}\;\;\;\phi\mapsto\W\phi=\Psi
\end{equation}
is a norm-preserving linear map (an isometry), because from (\ref{res}) we have
$$\|\W\phi\|^2=\|\Psi\|^2=\int_{\C}\abs{\Psi(z)}^2d\mu(z)=\|\phi\|^2.$$
 The range of this isometry, $\W\IH=\IH_{K}\subset\WH$ is a closed subspace of $\WH$. Let
 \begin{equation}\label{rep}
K_{T}(z,z')=\langle T(z)\mid T(z')\rangle=\frac{1}{\sqrt{\N(|z|)\N(|z'|)}}\sum_{m=0}^{\infty}\frac{\overline{T^m(z)}T^m(z')}{\rho(m)},
\end{equation}
which satisfies
\begin{equation}\label{cond}
\Psi(z)=\int_{\C}K_{T}(z,z')\Psi(z')d\mu(z'),
\end{equation}
a reproducing property satisfied by any vector $\Psi\in\IH_{K}$ with the reproducing kernel $K_{T}(z,z')$. The function $K_{T}$ satisfies
\begin{eqnarray}
&&K_{T}(z,z')=\overline{K_{T}(z',z)}\\
&&K_{T}(z,z)>0\\
&&\int_{\mathbb C}K_{T}(z,z'')K_{T}(z'',z')d\mu(z'')=K_{T}(z,z')\label{sq}.
\end{eqnarray}
The property (\ref{sq}) is called the square integrability property of $K_{T}$. Using (\ref{map2}) we can write
$$\left(\W\mid T(z')\rangle\right)(z)=\langle T(z)\mid T(z')\rangle=K_{T}(z,z').$$
Thus for a fixed $z'$ the function $z\mapsto K_{T}(z,z')$ is the image in $\IH_{K}$ of $\mid T(z')\rangle$ under the isometry $\W$. Since $\{\mid T(z)\rangle:z\in\C\}$ is overcomplete in $\IH$ and $\W$ is an isometry, the set of vectors $\{\W\mid T(z)\rangle:z\in\C\}$ is overcomplete in $\IH_{K}$, that is,
$$\{\eta_{z}=K_{T}(\cdot\;,z)\;:\;z\in\C\}$$
is overcomplete in $\IH_{K}$. Observe that $\eta_{z}$ are the same CS as $\mid T(z)\rangle$ but now written as vectors in the Hilbert space of functions $\IH_{K}$. From the above argument it is clear that $\IH_{K}$ is a reproducing kernel Hilbert space of the reproducing kernel $K_{T}$. These results are the analogue of the harmonic oscillator canonical CS. For a detail discussion regarding the reproducing kernels and CS, \cite{key1} is an excellent reference.
\begin{example}
Let $T:\mathbb C\longrightarrow\mathbb C$ by $T(z)=z^{k}$, where $k\not=1$ is a fixed positive integer.
Then with $z=re^{i\theta}$
$$T^{m}(z)=z^{k^{m}}=r^{k^{m}}e^{ik^{m}\theta}\hspace{0.5cm}\text{and}\hspace{0.5cm}\overline{T^{m}(z)}=\overline{z}^{k^{m}}=r^{k^{m}}e^{-ik^{m}\theta}$$
Let $\rho(m)=\Gamma(k^{m}+1)$. Since $T^{m}(z)\overline{T^{m}(z)}=r^{2k^{m}}$, $\{\frac{r^{2k^{m}}}{\Gamma(k^{m}+1)}\}_{m=0}^{\infty}$ is a subsequence of the positive sequence $\{\frac{r^{2m}}{\Gamma(m+1)}\}_{m=0}^{\infty}$ and the sequence $\{\frac{r^{2m}}{\Gamma(m+1)}\}_{m=0}^{\infty}$ is summable, we have
$$\mathcal N(|z|)=\sum_{m=0}^{\infty}\frac{r^{2k^{m}}}{\Gamma(k^{m}+1)}<\infty.$$
Further,
let $d\mu(r,\theta)=\frac{\mathcal N(|z|)}{\pi}e^{-r^{2}}rdrd\theta$ then
\begin{eqnarray*}
\int_{\mathbb C}\frac{1}{\mathcal N(|z|)}T^{m}(z)\overline{T^{l}(z)}d\mu &=&\frac{1}{\pi}\int_{0}^{\infty}\int_{0}^{2\pi}r^{k^{m}}e^{ik^{m}\theta}r^{k^{l}}e^{-ik^{l}\theta}e^{-r^{2}}rdrd\theta\\
&=&\left\{\begin{array}{ccc}
0&\text{if}&m\not=l\\
\Gamma(k^{m}+1)&\text{if}&m=l
\end{array}\right.
\end{eqnarray*}
where we have used the identity
$$\Gamma(z)=\int_{0}^{\infty}e^{-t}t^{z-1}dt\hspace{0.5cm}\text{when}\hspace{0.5cm}\text{Re}z>0$$
and the fact that
$$\int_{0}^{2\pi}e^{i(k^{m}-k^{l})\theta}d\theta=\left\{\begin{array}{ccc}
0&\text{if}&m\not=l\\
2\pi&\text{if}&m=l
\end{array}\right.$$
Thus with $T(z)=z^{k}$ the collection of vectors
\begin{equation}\label{6}
\mid T(z)\rangle=\mathcal N(|z|)^{-\frac{1}{2}}\sum_{m=0}^{\infty}\frac{T^{m}(z)}{\sqrt{\Gamma(k^{m}+1)}}\phi_{m};\;\;z\in\C
\end{equation}
forms a set of CS.
Here $\mathcal N(|z|)$ doesn't wear a closed form. In this case the reproducing kernel (\ref{rep}) takes the form
$$K_{T}(z,z')=\frac{1}{\sqrt{\N(|z|)\N(|z'|)}}\sum_{m=0}^{\infty}\frac{(\overline{z}z')^{k^m}}{\Gamma(k^m+1)}.$$
Since
$$\langle\phi\mid T(z)\rangle=\N(|z|)^{-\frac{1}{2}}\sum_{m=0}^{\infty}\frac{\langle\phi\mid\phi_m\rangle}{\sqrt{\Gamma(k^m+1)}}z^{k^m}=\N(|z|)^{-\frac{1}{2}}f(z)$$
where $f$ is an analytic function of the complex variable $z$, the image of the isometry $\KW$ is a space of analytic functions in $z$ that are square integrable with respect to the measure $d\mu(r,\theta)$.\\
It is an analogous result of the canonical CS. However, the states (\ref{6}) cannot be transformed to the form of $(\ref{E})$. Further, for the states (\ref{E}) there is always an associated Lie algebra but to the states (\ref{6}) we cannot define such algebra in the way that has been used to define algebra for the states (\ref{E}). See section \ref{osc}.
\end{example}
\section{CS with higher functions}
In this section we discuss two classes of CS by taking $\Phi_m$ as an associated Laguerre function and the Bessel function of the first kind. That is
$$\Phi_m(r,\theta)=\left[L_{m}^{\alpha}(r)\right]^{\frac{1}{2}}e^{im\theta},\;\;\;\Phi_m(r,\theta)=\left[r^mJ_{m+\frac{1}{2}}(r)\right]^{\frac{1}{2}}e^{im\theta}$$
We also discuss a reproducing kernel Hilbert space to each class. The inserted term $e^{im\theta}$ in the Laguerre case plays an important role in getting a resolution of the identity. In the case of the Bessel function, the term $r^{\frac{m}{2}}e^{im\theta}$ plays important roles in both, the normalization and resolution of the identity.
\subsection{CS with Laguerre Polynomials}
The Laguerre polynomials are given by
\begin{equation}\label{Lag}
L^{\alpha}_{m}(x)=\frac{e^{x}x^{-\alpha}}{m!}\frac{d^{m}}{dx^{m}}(e^{-x}x^{m+\alpha}).
 \end{equation}
\begin{proposition}\label{prop1}
For $r\in[0,\infty)$ and $\theta\in[0,2\pi)$, the vectors
\begin{equation}
\mid r,\theta\rangle= [e^{r}r^{-\alpha}\Gamma(\alpha,r)]^{-\frac{1}{2}}\sum_{m=0}^{\infty}\frac{e^{im\theta}\left[L_{m}^{\alpha}(r)\right]^{\frac{1}{2}}}{\sqrt{m+1}}\phi_{m}\label{lag1}
\end{equation}
form a set of CS.
\end{proposition}
\begin{proof}
From \cite{key15} (see page 91, formula 4.24.5) we have
\begin{equation}
\sum_{m=0}^{\infty}\frac{L_{m}^{\alpha}(r)}{m+1}=e^{r}r^{-\alpha}\Gamma(\alpha,r),\label{lag3}
\end{equation}
where $0< r<\infty$ and $\alpha>-1$.
Thus we have
$$\langle r,\theta\mid r,\theta\rangle=[e^{r}r^{-\alpha}\Gamma(\alpha,r)]^{-1}\sum_{m=0}^{\infty}\frac{L_{m}^{\alpha}(r)}{m+1}=1.$$
On $\D=[0,\infty)\times[0,2\pi),$ let us take the measure,
$$d\mu(r,\theta)=\frac{\mathcal N(r)}{2\pi}\lambda(r)drd\theta,$$
where $$\lambda(r)=\frac{ r^{\beta-1}e^{-r}}{\Gamma(\beta)}$$
 is an auxiliary density and $\beta$ is a positive constant such that $\alpha-\beta=1$.
For a resolution of the identity, consider
\begin{eqnarray*}
&&\int_{0}^{\infty}\int_{0}^{2\pi}\mid r,\theta\rangle\langle r,\theta\mid d\mu(r,\theta)\\
&=&\sum_{m=0}^{\infty}\frac{1}{(m+1)\Gamma(\beta)}\int_{0}^{\infty}r^{\beta-1}e^{-r}L_{m}^{\alpha}(r)dr\mid\phi_{m}\rangle\langle\phi_{m}\mid.
\end{eqnarray*}
Thus the resolution of the identity demands,
\begin{equation}
\int_{0}^{\infty}r^{\beta-1}e^{-r}L_{m}^{\alpha}(r)dr=(m+1)\Gamma(\beta).\label{lag4}
\end{equation}
 Now we use the following transform from \cite{key3},Vol-2, page-292 (1).
\begin{equation}
\int_{0}^{\infty}x^{\beta-1}e^{- x}L_{m}^{\alpha}(x)dx=\frac{\Gamma(\alpha-\beta+m+1)\Gamma(\beta)}{m!\Gamma(\alpha-\beta+1)}\label{lag5}
\end{equation}
where $\beta>0$. Since $\alpha-\beta=1$ from (\ref{lag4}) and (\ref{lag5}) we get a resolution of the identity.
\end{proof}
Using the definition of the associated Laguerre function (\ref{Lag}) one can rewrite the CS as follows:
\begin{equation}\label{deri}
\mid r,\theta\rangle=\Gamma(\alpha,r)^{-\frac{1}{2}}\sum_{m=0}^{\infty}\frac{e^{im\theta}\left[F_{\alpha}^{(m)}(r)\right]^{\frac{1}{2}}}{\sqrt{(m+1)!}}\phi_m
\end{equation}
where $F_{\alpha}(r)=e^{-r}r^{m+\alpha}$ and $F_{\alpha}^{(m)}(r)$ is the $m^{\text{th}}$ derivative of it. In this case the associated reproducing kernel takes the form
$$K_{F}(r,\theta,r',\theta')=[\Gamma(\alpha,r)\Gamma(\alpha,r')]^{-\frac{1}{2}}\sum_{m=0}^{\infty}\frac{\sqrt{F_{\alpha}^{(m)}(r)F_{\alpha}^{(m)}(r')}}{(m+1)!}e^{im(\theta'-\theta)}.$$
and the square integrability condition reads
$$\int_{0}^{\infty}\int_{0}^{2\pi}K_F(r,\theta,r'',\theta'')K_F(r'',\theta'',r',\theta')d\mu(r'',\theta'')=K_F(r,\theta,r',\theta').$$
The image of the isometry $\KW$ becomes,
$$(\KW\phi)(r,\theta)=\Gamma(\alpha,r)^{-\frac{1}{2}}\sum_{m=0}^{\infty}\frac{\langle\phi\mid\phi_m\rangle}{\sqrt{(m+1)!}}e^{im\theta}\left[F_{\alpha}^{(m)}(r)\right]^{\frac{1}{2}}=\Gamma(\alpha,r)^{-\frac{1}{2}}f(r,\theta).$$
 Now, in accordance with (\ref{lag1}), let us see the normalization factor and the measure for some special values of $\alpha$ and $\beta$.
\begin{enumerate}
\item[(i)]
For $\alpha=2$ and $\beta=1$,
$$\mathcal N(r)=\frac{1+r}{r^2}\;\;\;\text{and}\;\;\;\lambda(r)=e^{-r};\;\;\;r>0.$$
$\mathcal N(r)$ is singular at $r=0$.
\item[(ii)]
For $\alpha=3$ and $\beta=2$,
$$\mathcal N(r)=\frac{r^2+2r+2}{r^3}\;\;\;\text{and}\;\;\;\lambda(r)=re^{-r};\;\;\;r>0.$$
$\mathcal N(r)$ is singular at $r=0$.
\item[(iii)]
For $\alpha=4$ and $\beta=3$,
$$\mathcal N(r)=\frac{r^3+3r^2+6r+6}{r^4}\;\;\;\text{and}\;\;\;\lambda(r)=\frac{r^2e^{-r}}{2};\;\;\;r>0.$$
$\mathcal N(r)$ is singular at $r=0$.
\end{enumerate}

\subsection{CS with Bessel function}
Here we discuss a set of CS with Bessel functions of the first kind, which is defined as
\begin{equation}\label{J}
J_{\nu}(z)=\sum_{m=0}^{\infty}\frac{(-1)^m\left(\frac{z}{2}\right)^{\nu+2m}}{m!\Gamma(\nu+m+1)}
\end{equation}
\begin{proposition}\label{prop2}
For $r\in[0,\infty)$ and $\theta\in[0,2\pi)$  the set of vectors
\begin{equation}
\mid r,\theta\rangle=\left[I_{\frac{1}{2}}(r)\right]^{-\frac{1}{2}}\sum_{m=0}^{\infty}\left[\frac{r^{m}J_{m+\frac{1}{2}}(r)}{m!}\right]^{\frac{1}{2}}e^{im\theta}\phi_{m}\label{b1},
\end{equation}
where $I_{\nu}(r)$ is the order $\nu$ modified Bessel function of the second kind, form a set of CS.
\end{proposition}
\begin{proof}
We have
$$\langle r,\theta\mid r,\theta\rangle=[I_{\frac{1}{2}}(r)]^{-1}\sum_{m=0}^{\infty}\frac{r^{m}J_{m+\frac{1}{2}}(r)}{m!}=1,$$
where we have used the identity (see \cite{key16}, page 296)
$$\sum_{m=0}^{\infty}\frac{x^{m}}{\Gamma(m+1)}J_{n+m}(x)=I_{n}(x)$$
with $n=\frac{1}{2}$. Since $I_{\frac{1}{2}}(r),$ is a sharply increasing positive function, the measure, on $\D=[0,\infty)\times[0,2\pi)$,
$$d\mu(r,\theta)=\sqrt{\frac{r}{2\pi}}e^{-r}I_{\frac{1}{2}}(r)drd\theta$$
is  positive. With this measure let us see a resolution of the identity.
\begin{eqnarray*}
&&\int_{0}^{\infty}\int_{0}^{2\pi}\mid r,\theta\rangle\langle r,\theta\mid d\mu(r,\theta)\\
&=&\sum_{m=0}^{\infty}\sum_{l=0}^{\infty}\frac{\mid\phi_{m}\rangle\langle\phi_{l}\mid}{\sqrt{m!l!}}\int_{0}^{\infty}\int_{0}^{2\pi}\frac{e^{i(m-l)\theta}}{\mathcal N(r)}\sqrt{r^{m+l}J_{m+\frac{1}{2}}(r)J_{l+\frac{1}{2}}(r)}\sqrt{\frac{r}{2\pi}}e^{-r}I_{\frac{1}{2}}(r)drd\theta\\
&=&\sum_{m=0}^{\infty}\frac{\sqrt{2\pi}}{m!}\int_{0}^{\infty}r^{m+\frac{1}{2}}J_{m+\frac{1}{2}}(r)e^{-r}dr\mid\phi_{m}\rangle\langle\phi_{m}\mid\\
&=&\sum_{m=0}^{\infty}\mid\phi_{m}\rangle\langle\phi_{m}\mid=I,
\end{eqnarray*}
where we have used the identity (see \cite{key16}, page 260, formula 6.55)
$$\int_{0}^{\infty}e^{-ax}x^{p}J_{p}(bx)dx=\frac{(2b)^{p}\Gamma(p+\frac{1}{2})}{\sqrt{\pi}(a^{2}+b^{2})^{p+\frac{1}{2}}};\;\;\;\;\;p>-\frac{1}{2},\;a,b>0$$
with $b=1,\;a=1$ and $p=m+\frac{1}{2}.$ Thus the states in (\ref{b1}) form a set of CS.
\end{proof}
Here again the image of the isometry and a reproducing kernel can be obtained as in the case of Laguerre polynomials.

\section{The oscillator algebra}\label{osc}
For the states in (\ref{E}), there is a natural way of defining the annihilation, creation and number operators. These three operators are denoted respectively by $a,a^{\dagger}$ and $N$. Let $x_{m}=\frac{\rho(m)}{\rho(m-1)}$ and $x_{0}!=1$ then $\rho_{m}=x_{m}!$, the generalized factorial. For the basis vectors $\{\phi_{m}\}_{m=0}^{\infty}$, these operators are defined as
\begin{equation}
a\phi_{m}=\sqrt{x_{m}}\phi_{m-1},\;\;\; a^{\dagger}\phi_{m}=\sqrt{x_{m+1}}\phi_{m+1},\;\;\;\;\text{and}\;\;\;N\phi_{m}=x_{m}\phi_{m}\label{a1},
\end{equation}
where $a^{\dagger}$ is the adjoint of $a$ and $N=a^{\dagger}a$. For a detail explanation see \cite{key1}. Under this definition, the commutators take the form
\begin{eqnarray}
&&[a,a^{\dagger}]\phi_m=(x_{m+1}-x_{m})\phi_m,\\
&&[N,a^{\dagger}]\phi_m=(x_{m+1}-x_{m})\phi_{m+1}\;\;\text{and}\\
&&[N,a]\phi_m=(x_{m-1}-x_{m})\phi_{m-1}.\label{a2}
\end{eqnarray}
These three operators, under the commutator bracket, generate a Lie algebra $\mathfrak U_{\text{osc}}$ called a generalized oscillator algebra.
Under this definition the states in (\ref{E}) become the eigenstate of $a$, i.e, $$a\mid z\rangle=z\mid z\rangle.$$
Since $a^{\dagger}$ is the adjoint of $a$ and $N=a^{\dagger}a$, if we define the annihilation operator for a set of CS we can obtain the operators $a^{\dagger}$ and $N$. To define an annihilation operator, in principle, for a set of CS we require
\begin{equation}
a\phi_{m}=g(m)\phi_{m-1},\;a\phi_{0}=0,\;\text{and}\; a\mid\cdot\rangle=f(\cdot)\mid\cdot\rangle,\label{ann1}
\end{equation}
where $g(m)$ depends only on $m$ and $f(\cdot)$ doesn't depend on $m$ but depends only on the labeling parameter of CS.
In the following we will show that for the states in (\ref{6}), (\ref{lag1}) and (\ref{b1}) an annihilation operator cannot be defined to satisfy (\ref{ann1}). However, we will build CS agreeing with (\ref{1}) to have a Lie algebra in the following section. For the states in (\ref{6}), suppose there exists an annihilation operator to satisfy (\ref{ann1}). Then
\begin{eqnarray}
a\mid T(z)\rangle&=&f(z)\mid T(z)\rangle\;\;\text{and}\label{ann2}\\
a\phi_{m}&=&g(m)\phi_{m-1}\label{ann3}
\end{eqnarray}
By (\ref{ann3}) we get,
$$a\mid T(z)\rangle=\mathcal N(|z|)^{-\frac{1}{2}}\sum_{m=1}^{\infty}\frac{z^{k^{m}}g(m)}{\sqrt{\rho(m)}}\phi_{m-1}
=\mathcal N(|z|)^{-\frac{1}{2}}\sum_{m=0}^{\infty}\frac{z^{k^{m+1}}g(m+1)}{\sqrt{\rho(m+1)}}\phi_{m}.$$
Now from (\ref{ann2}) we get
$$\mathcal N(|z|)^{-\frac{1}{2}}\sum_{m=0}^{\infty}\frac{z^{k^{m+1}}g(m+1)}{\sqrt{\rho(m+1)}}\phi_{m}
=\mathcal N(|z|)^{-\frac{1}{2}}\sum_{m=0}^{\infty}\frac{z^{k^{m}}f(z)}{\sqrt{\rho(m)}}\phi_{m}.$$
Since $\{\phi_{m}\}$ is an orthonormal basis we get
$$\frac{z^{k^{m+1}}g(m+1)}{\sqrt{\rho(m+1)}}=\frac{z^{k^{m}}f(z)}{\sqrt{\rho(m)}}\;\;\;\text{for all}\;m.$$
Thus
\begin{equation}
f(z)=z^{k^{m+1}-k^{m}}g(m+1)\sqrt{\frac{\rho(m)}{\rho(m+1)}}\label{ann6}.
\end{equation}
Since $f(z)$ is independent of $m$ we must have
\begin{eqnarray}
g(m+1)\sqrt{\frac{\rho(m)}{\rho(m+1)}}&=&C_{1}\;\;\;\text{and}\label{ann4}\\
k^{m+1}-k^{m}=k^{m}(k-1)&=&C_{2}\label{ann5},
\end{eqnarray}
where $C_{1}$ and $C_{2}$ are constants. $\rho(m)$ and $g(m)$ can be chosen to satisfy (\ref{ann4}) but (\ref{ann5}) can be satisfied only for $k=1$. In our construction we assumed $k\not=1$. Thus for the states in (\ref{6}) an annihilation operator cannot be defined to satisfy (\ref{ann1}). For the general case (\ref{5}), $f(z)$ becomes
$$f(z)=\frac{T^{m+1}(z)}{T^{m}(z)}g(m+1)\sqrt{\frac{\rho(m)}{\rho(m+1)}}.$$
Thus an annihilation operator can only be defined if
$$\frac{T^{m+1}(z)}{T^{m}(z)}=h(z),$$
a function of $z$ only. In general, it is not a possibility. For the states in (\ref{lag1}), $f(z)$ takes the form
$$f(z)=\sqrt{\frac{{L}_{m+1}^{\alpha}(r)}{{L}_{m}^{\alpha}(r)}}g(m+1)\sqrt{\frac{\rho(m)}{\rho(m+1)}}.$$
Again, an annihilation operator can only be defined if
$$\sqrt{\frac{{L}_{m+1}^{\alpha}(r)}{{L}_{m}^{\alpha}(r)}}=k(r),$$
a function independent of $m$, which is not the case. Therefore, for the states (\ref{lag1}) there cannot be an annihilation operator in the form (\ref{ann1}). A similar argument applies to the CS in (\ref{b1}).
\section{CS with elementary functions}\label{Eli}
In this section we present classes of CS by taking $\Phi_m$ as some elementary functions of type $\Phi_m(z)=F(z)^{m}$, the $m^{\text{th}}$ power of a function $F(z)$. In these cases, as we mentioned earlier, we will also generate a Lie algebra structure associated to CS. 

\subsection{On an arbitrary disc:}
Here we discuss a set of CS with $$\Phi_m(r,\theta)=e^{im\theta}(y-r)^m$$ where $y$ is a fixed positive real number, $r\in (0,y)$ and $\theta\in[0,2\pi)$. Let $$\D_{y}=\{z=(y-r)e^{i\theta}\;:\;r\in (0,y),\theta\in[0,2\pi)\},$$ a disc of radius $y$. Now we define a set of CS on $\D_{y}$.
\begin{proposition}\label{prop3}
For $z\in\D_{y}$, the set of vectors
\begin{equation}
\mid z\rangle=\mathcal N(r,y)^{-\frac{1}{2}}\sum_{m=0}^{\infty}\left[\frac{(\nu+1)_{2m}}{(2m)!y^{2m+\nu}}\right]^{\frac{1}{2}}z^{m}\phi_{m}\label{18}
\end{equation}
form a set of CS, where
$$\mathcal N(r,y)=\frac{\Gamma(1+\nu)}{\Gamma(\nu)y^{\nu}}\sqrt{\frac{\pi(1-r)}{y}}\left(1-\frac{(1-r)^{2}}{y^{2}}\right)^{-\frac{\nu+2}{2}}P_{-\frac{\nu+2}{2}}^{\frac{1}{2}}\left(\frac{y^{2}+(1-r)^{2}}{y^{2}-(1-r)^{2}}\right),$$
where $P_{m}^{l}(x)$ is the associated Legendre polynomials, and
$$(a)_m=\frac{\Gamma(a+m)}{\Gamma(a)}$$
the Pochhammer symbol.
\end{proposition}
\begin{proof}
In order to obtain the normalization and a resolution of the identity, we use the following integral transform:
From \cite{key3}, Vol.2, page-185-(7) the following transform
\begin{equation}
\int_{0}^{y}x^{\nu-1}(y-x)^{\mu-1}dx=\frac{\Gamma(\mu)\Gamma(\nu)}{\Gamma(\mu+\nu)}y^{\mu+\nu-1}\label{15}
\end{equation}
is valid when  $\mu,\nu>0$ and $y>0$. Since we require $\langle z\mid z\rangle=1$ and
$$\langle z\mid z\rangle=\mathcal N(r,y)^{-1}\sum_{m=0}^{\infty}\frac{(\nu+1)_{2m}}{(2m)!y^{2m+\nu}}(y-r)^{2m}$$
the normalization factor takes the form
\begin{eqnarray*}
\mathcal N(r,y)&=&\sum_{m=0}^{\infty}\frac{(\nu+1)_{2m}}{(2m)!y^{2m+\nu}}(y-r)^{2m}\\
&=&\frac{\Gamma(1+\nu)}{\Gamma(\nu)y^{\nu}}\sqrt{\frac{\pi(1-r)}{y}}\left(1-\frac{(1-r)^{2}}{y^{2}}\right)^{-\frac{\nu+2}{2}}P_{-\frac{\nu+2}{2}}^{\frac{1}{2}}\left(\frac{y^{2}+(1-r)^{2}}{y^{2}-(1-r)^{2}}\right).
\end{eqnarray*}
Since
$$\lim_{m\rightarrow\infty}{\frac{(y-r)^{2m+2}\rho(m)}{(y-r)^{2m}\rho(m+1)}}=\frac{y^{2}-2yr+r^{2}}{y^{2}}$$
the series converges on $[0,y)$.
For the resolution of the identity, let us take a measure on $\D_{y}$ as
$$d\kappa(r,\theta)=d\omega(r)d\theta.$$
Consider
\begin{eqnarray*}
\int_{\D_{y}}\mid z\rangle\langle z\mid d\kappa(r,\theta)&=&\int_{0}^{y}\int_{0}^{2\pi}\mid r,\theta\rangle\langle r,\theta\mid d\omega(r)d\theta\\
&=&\sum_{m=0}^{\infty}2\pi \frac{(\nu+1)_{2m}}{(2m)!y^{2m+\nu}}\int_{0}^{y}\frac{(y-r)^{2m}}{\mathcal N(r)}d\omega(r)\mid\phi_{m}\rangle\langle\phi_{m}\mid.
\end{eqnarray*}
 Let us take $d\omega(r)$ as
$$d\omega(r)=\frac{\mathcal N(r)}{2\pi}\lambda(r)dr,$$ where $\lambda(r)=r^{\nu-1}$ is an auxiliary density. Thus to get
$$\int_{\D_{y}}\mid z\rangle\langle z\mid d\kappa(r,\theta)=I$$
we must have
$$\frac{(\nu+1)_{2m}}{(2m)!y^{2m+\nu}}\int_{0}^{y}r^{\nu-1}(y-r)^{2m}dr=1$$
which is true  by (\ref{15}) with $\mu-1=2m$. Thus the set of vectors $\{\mid z\rangle\;:\;z\in\D_y\}$ forms a set of CS.
\end{proof}
Now for a reproducing kernel, let us take
$$K(\overline{z},z')=\langle \overline{z}\mid z'\rangle.$$
Through a small calculation we arrive, 
$$K(\overline{z},z')=\sqrt{\frac{\pi}{\N(r,y)\N(r',y)}}\frac{\Gamma(\nu+1)(zz')^{\frac{1}{4}}y^{-\frac{3}{2}}}{(y^2-zz')^{\frac{\nu+2}{2}}}P_{-\frac{(\nu+2)}{2}}^{\frac{1}{2}}\left(\frac{zz'+y^2}{y^2-zz'}\right)$$
and it satisfies the square integrability property
$$\int_{\mathcal D_{y}}K(\overline{z},z'')K(\overline{z''},z')d\kappa(r,\theta)=K(\overline{z},z').$$
The isometric image takes the form
$$(\KW\phi)(z)=\langle\phi\mid z\rangle=\N(r,y)^{-\frac{1}{2}}\sum_{m=0}^{\infty}\langle\phi\mid\phi_m\rangle\left[\frac{(\nu+1)_{2m}}{(2m)!y^{2m+\nu}}\right]^{\frac{1}{2}}z^m=\N(r,y)^{-\frac{1}{2}}f(z),$$
where $f$ is an analytic function in the variable $z$ on $\D_{y}$. It is interesting to notice that by changing $y$ we can change the domain of interest as we please.
Let us see some special values. For $y=1$ and $\nu=1$ we get
$$\rho(m)=\frac{1}{2m+2}\;\;\;\;\;\mathcal N(r)=\frac{\sqrt{\pi(1-r)}}{(2r-r^{2})^{2}}P_{\frac{1}{2}}^{\frac{1}{2}}\left(\frac{r^{2}-2r+2}{2r-r^2}\right),\;\;\lambda(r)=1$$
where $\mathcal N(r)$ is positive and has a singularity at $r=0$.

For $y=2$ and $\nu=2$ we have
$$\rho(m)=\frac{2^{2m+2}}{(2m+2)(2m+3)}\;\;\;\;\;\mathcal N(r)=\frac{4\sqrt{\pi(2-r)}}{\sqrt{2}r^{2}(4-r)^{2}}P_{1}^{\frac{1}{2}}\left(\frac{r^{2}-4r+8}{4r-r^2}\right),\;\;\lambda(r)=r,$$
again  $\mathcal N(r)$ is positive and has a singularity at $r=0$. In a similar way by changing $y$ and $\nu$ we can generate several classes of CS of type (\ref{18}).\\

 Further, an oscillator algebra can be defined, for example for the case $y=1$ and $\nu=1$ we have 
$$x_{m}=\frac{m-1}{m+1}$$
in (\ref{a1}). Thus by the discussion of section \ref{osc} we  have a Lie algebra, $\mathfrak U_{\text{osc}}$ associated to the CS but it cannot be identified with a classical Lie algebra. In these cases, we also have 
$$a \mid z\rangle=z\mid z\rangle,$$
that is, the CS is an eigenstate of the annihilation operator, $a$.
\subsection{Another class of CS on the unit disc}
Here we discuss another class of CS on the unit disc with a different choice of $\Phi_m$. These CS are in analogy with the Barut-Girardello CS in the sense that these CS can be realized as the eigenstates of $K_{-}$, one of the generators of the classical $su(1,1)$ Lie algebra. For the details of $su(1,1)$, for example, see \cite{key1}.   Let
$$\D=\{(r,\theta)\;:\;0\leq r<1,\;0\leq\theta<2\pi\}.$$
Let us consider the sequence of functions
$$\Phi_{m}(r,\theta)=e^{im\theta}[\log{r}]^m,\;\;\;m=0,1,2,\dots$$

\begin{proposition}\label{prop4}
For $(r,\theta)\in \D$ and $z=e^{i\theta}\log{r}$ the vectors
\begin{equation}
\mid z\rangle=[\cosh{(\text{csgn}{(\log{r})}\log{r})}]^{-\frac{1}{2}}\sum_{m=0}^{\infty}\frac{z^m}{\sqrt{(2m)!}}\phi_{m},\;\;\;r>0\label{27},
\end{equation}
where
$$\text{csgn}(r)=\left\{\begin{array}{ccc}
1&\text{if}&r>0\\
0&\text{if}&r=0\\
\end{array}\right.$$
form a set of CS.
\end{proposition}
\begin{proof}
Let us first see the normalization
\begin{equation}
\langle z\mid z\rangle=[\cosh{(\text{csgn}{(\log{r})}\log{r})}]^{-1}\sum_{m=0}^{\infty}\frac{[\log{r}]^{2m}}{(2m)!}=1,\;\;\;0<r<1\label{28}
\end{equation}
Note that the normalization factor has a singularity at $r=0$.
On $\D$ we take the measure,
$$d\kappa(r,\theta)=\frac{\mathcal N(r)}{2\pi r^{2}}d\theta dr,$$
which is singular at $r=0$. Now for the resolution of the identity, consider
\begin{eqnarray*}
\int_{\D}\mid z\rangle\langle z\mid d\kappa(r,\theta)
&=&\sum_{m=0}^{\infty}\frac{1}{\rho(m)}\int_{0}^{1}[(\log{r})]^{2m}r^{-2}dr\mid\phi_{m}\rangle\langle\phi_{m}\mid\\
&=&\sum_{m=0}^{\infty}\mid\phi_{m}\rangle\langle\phi_{m}\mid=I,
\end{eqnarray*}
where we have used the integral transform (\cite{key3},Vol1,page 315, (14)) with $s=-1$ and $\nu=2m+1$. The transform
$$\int_{0}^{1}(\log{x})^{\nu-1}x^{s-1}dx=(-s)^{-\nu}\Gamma(\nu)$$
is valid when $0<x<1$, $\nu>0$ and $s<0$. 
\end{proof}
In this case, the reproducing kernel takes the form
$$K(\overline{z},z')=[\cosh{(\text{csgn}{(\log{r})}\log{r})}\cosh{(\text{csgn}{(\log{r'})}\log{r'})}]^{-\frac{1}{2}}\cosh{\sqrt{zz'}}$$
and a reproducing condition similar to the previous case can be written. The isometric image is
\begin{eqnarray*}
(\KW\phi)(r,\theta)&=&[\cosh{(\text{csgn}{(\log{r})}\log{r})}]^{-\frac{1}{2}}\sum_{m=0}^{\infty}\frac{\langle\phi\mid\phi_m\rangle}{\sqrt{(2m)!}}z^m\\
&=&[\cosh{(\text{csgn}{(\log{r})}\log{r})}]^{-\frac{1}{2}}f(z).
\end{eqnarray*}
where $f(z)$ is an analytic function of $z$ in $\D$.
Let us see the oscillator algebra $\mathfrak U_{\text{osc}}$ associated to this class of CS.
Since 
$$x_{m}=\frac{\Gamma(2m+1)}{\Gamma(2m-2)}=2m(2m-1)$$
from (\ref{a1})  we get
\begin{eqnarray*}
a\phi_m&=&\sqrt{2m(2m-1)}\phi_{m-1}=2\sqrt{m(m-\frac{1}{2})}\phi_{m-1}\\
a^{\dagger}\phi_m&=&\sqrt{2(m+1)(2m+1)}\phi_{m+1}=2\sqrt{(m+1)(m+\frac{1}{2})}\phi_{m+1}\\
N\phi_m&=&2m(2m-1)\phi_m=4m(m-\frac{1}{2})\phi_m
\end{eqnarray*}
Now let us define a new set of operators
\begin{eqnarray*}
\PA\phi_m&=&\frac{1}{2}a\phi_{m}=\sqrt{m(m-\frac{1}{2})}\phi_{m-1}\\
\PA^{\dagger}\phi_m&=&\frac{1}{2}a^{\dagger}\phi_{m}=\sqrt{(m+1)(m+\frac{1}{2})}\phi_{m+1}\\
\mathfrak{N}\phi_m&=&(m+\frac{1}{4})\phi_m
\end{eqnarray*}
Note that in this case $\mathfrak{N}\not=\PA^{\dagger}\PA$. By a direct calculation it can be seen that these new operators satisfy the commutation relations,
$$[\PA,\PA^{\dagger}]=2\mathfrak{N},\;\;\;[\mathfrak{N},\PA]=-\PA,\;\;\;[\mathfrak{N},\PA^{\dagger}]=\PA^{\dagger}.$$
These are the exact commutation relations satisfied by the generators $K_{-},K_{+},K_{3}$ of the classical Lie algebra $su(1,1)$ of the classical group $SU(1,1)$.
  Thus, in this sense, we have $\mathfrak U_{\text{osc}}=su(1,1)$. We also have 
$$a\mid z\rangle=z\mid z\rangle\;\;\;\text{and}\;\;\;\PA\mid z\rangle=\frac{z}{2}\mid z\rangle.$$
From this last relation, constructed CS can also be considered as CS of Barut-Girardello type.
\section{Remarks and conclusion}
The CS presented in section 5 can serve as a quantum tool via the associated oscillator algebras. The states of sections 2 and 3 do not accommodate oscillator algebra in the usual way. As of now, to our knowledge, a physical application of these CS is only conjectural. However, mathematically these vectors form a set of CS and associate a reproducing kernel and a reproducing kernel Hilbert space.

\bigskip

The CS presented in section 2 may not easily be extended to complicated functions. For example, if we take $T(z)=z^{2}+c$, where $c$ is a real or complex constant, then we do not have a closed form for $T^{m}(z)$ which could severely restrict our ability of getting the normalization and a resolution of the identity. The states presented in sections 3 may be extended to other higher functions. The states similar to the ones presented in section 5 can easily be manipulated from integral transforms or by other means. For the CS of section 2, if we restrict the parameter $z$ to be in the Julia set of $T$ we can have frames on fractals. Some results supporting the latter claim will be presented elsewhere.



\begin{thebibliography}{XXXX}
\bibitem[1]{key1} Ali, S.T., Antoine, J-P., Gazeau, J.P. {\em Coherent States, Wavelets and Their Generalizations}, Springer, New York (2000).
\bibitem[2]{Antoine} Antoine, J-P., Gazeau, J-P., Monceau, P., Klauder, J.R., Penson, K.A., {\em Temporally stable coherent states for infinite well and P\"oschl-Teller potentials}, J.Math.Phys. {\bf 42} (2001), 2349-2387.
\bibitem[3]{key17} Borzov, V.V., {\it Orthogonal polynomials and generalized oscillator algebras}, Integral Transforms and Special Functions {\bf 12} (2001), 115-138.
\bibitem[4]{key18} Borzov, V.V., Damaskinsky, E.V. and Yegorov, S.B., {\it Some remarks on the representations of the generalized deformed oscillator algebra}, preprint q-alg/9509022 v1; Zap. Nauch. Seminarov. LOMI {\bf 245}, 80-106 (1997) (in Russian).
\bibitem[5]{Du} Daubechies Ingrid, {\em Painless nonorthogonal expansions}, J.Math.Phys. {\bf 27} (1986), 1271-1282.
\bibitem[6]{key3} Erd\'elyi, A.,  Magnus, W.,  Oberhettinger, F.,  Tricomi, F.G., {\it Tables of integral transforms,} McGraw-Hill,  New York (1953).
\bibitem[7]{GK} Gazeau, J-P., Klauder, J.R., {\em  Coherent states for systems with discrete and continuous spectrum}, J.Phys.A: Math.Gen. {\bf 32} (1999), 123-132.
\bibitem[8]{key2} Klauder, J.R, Skagerstam, B.S,  {\em Coherent States, Applications in Physics and Mathematical Physics}, World Scientific, Singapore, (1985).
\bibitem[9]{key7} Klauder, J.R., Penson, K.A., Sixdeniers, J-M., {\em Constructing Coherent States through solutions of Steieljes and Hausdorff moment problems}, Phys. Rev. {\bf{A 64}} (2001), 013817.
\bibitem[10]{key16} Larry, C.Andrewws, {\em Special functions of mathematics for engineers}, 2nd ed., McGraw-Hill, New York (1992).
\bibitem[11]{key15} Lebedev, N.N., {\em Special functions and their applications}, Prentice-Hall, Englewood Cliffs, N.J. (1965).
\bibitem[12]{key4} P\'er\'elemov, A.M., {\em Generalized Coherent States and Their Applications}, Springer-Verlag, Berlin, (1986).
\bibitem[13]{key9}Thirulogasanthar, K., Twareque Ali, S., {\it A class of vector coherent states defined over matrix domains}, To appear, J.Math.Phys.

\end{thebibliography}
                      \end{document}